\documentclass[twocolumn,english,aps,prd,nofootinbib,hidelinks]{revtex4-2}
\usepackage{lmodern}
\usepackage[T1]{fontenc}
\usepackage[latin9]{inputenc}
\usepackage{xcolor}
\usepackage{babel}
\usepackage{varioref}
\usepackage{amsmath}
\usepackage{amssymb}
\usepackage{graphicx}
\usepackage[unicode=true]
{hyperref}

\makeatletter

\providecommand{\tabularnewline}{\\}

\usepackage{enumerate}
\usepackage{academicons}
\definecolor{orcidlogocol}{HTML}{A6CE39}

\makeatother

\begin{document}
\title{Revisiting Weak Energy Condition and wormholes in Brans-Dicke gravity}
\author{Hoang Ky Nguyen$\,$}
\email[\ \ ]{hoang.nguyen@ubbcluj.ro}

\affiliation{Department of Physics, Babe\c{s}--Bolyai University, Cluj-Napoca
400084, Romania}
\author{Mustapha Azreg-A\"inou$\,$}
\email[\ \ ]{azreg@baskent.edu.tr}

\affiliation{Ba\c{s}kent University, Engineering Faculty, Ba\u{g}lica Campus,
06790-Ankara, Turkey}
\date{September 1, 2024}
\begin{abstract}
\vskip2pt It is known that the formation of a wormhole typically
involves a violation of the Weak Energy Condition (WEC), but the reverse
is not necessarily true. In the context of Brans-Dicke gravity, the
\emph{generalized} Campanelli-Lousto solution, which we shall unveil
in this paper, demonstrates a WEC violation that coincides with the
appearance of \emph{unbounded} sheets of spacetime within the region
where $0<r<r_{\text{s}}$. The emergence of a wormhole in the region
where $r>r_{\text{s}}$ is thus only an indirect consequence of the
WEC violation. Whereas the two regions, $0<r<r_{\text{s}}$ and $r>r_{\text{s}}$,
in general are disconnected by \emph{physical} singularities at $r=r_{\text{s}}$,
they are both part of the same \emph{mathematical} solution, and their
behavior can provide insights into the WEC, which is a \emph{mathematical}
property of the solution. Furthermore, we utilize the generalized
Campanelli-Lousto solution to construct a Kruskal-Szekeres diagram,
which exhibits a ``gulf'' sandwiched between the four quadrants in
the diagram, a novel feature in Brans-Dicke gravity. Overall, our
findings shed new light onto a complex interplay between the WEC and
wormholes in the Brans-Dicke theory.
\end{abstract}
\maketitle

\section{\label{sec:Introduction}Introduction}

Wormholes are hypothetical spacetime structures that may interact
with ordinary matter and thus can be observed and even distinguished
from black holes \citep{Darmour,Bambi,Azreg,Dzhunushaliev,Cardoso,Konoplya,Nandi,Bueno,Cramer,Nedkova,Harko,Deligianni,Falco,p4,p5}.

To maintain a wormhole, it is often required to violate the Weak Energy
Condition (WEC) \citep{MorrisThorne-1988-1,MorrisThorne-1988-2,Kar-1994,Kar-2016,Visser-1989}.
In its geometric form, the WEC requires that for every future-pointing
timelike vector $t^{\mu}$, the inequality
\begin{equation}
G_{\mu\nu}\,t^{\mu}t^{\nu}\geq 0,
\end{equation}
must hold, where $G_{\mu\nu}$ denotes the Einstein tensor. Specifically,
for the vector $t^{\mu}=(1,0,0,0)$, the WEC yields $G_{00}\ge0$,
meaning positive energy density everywhere in any frame of reference.
In general relativity (GR), maintaining a wormhole requires the presence
of exotic matter, which possesses negative energy density to violate
the WEC. However, in generalized or modified theories of GR, the WEC
can be violated through the introduction of additional terms or corrections
in the gravitational sector. These modifications can effectively simulate
the presence of exotic matter without being genuinely exotic. In these
scenarios, the \emph{vacuum} field equation involving the metric $g_{\mu\nu}$
becomes more complex. Nonetheless, it can often be rearranged into
the form $G_{\mu\nu}=\frac{8\pi G}{c^{4}}T_{\mu\nu\text{(eff)}}$ where $T_{\mu\nu\text{(eff)}}$
includes the corrections to the gravitational sector and it acts as a `surrogate' or `effective' source of exotic matter (SET), enabling the violation of the WEC which takes the form,
\begin{equation}\label{WEC}
T_{\mu\nu\text{(eff)}}\,t^{\mu}t^{\nu}=\frac{c^4}{8\pi G} G_{\mu\nu}\,t^{\mu}t^{\nu}\geq 0,	
\end{equation}
and maintaining wormholes for modified gravity. We emphasize that we are considering Einstein theory of gravity as a reference theory and that the violation of the WEC is regarded with respect to his theory when all extra geometric terms are grouped into an effective SET.

An example of a theory that supports a wormhole is the Brans-Dicke
(BD) action. In \citep{Agnese-1995} Agnese and La Camera (ALC) used the
Campanelli-Lousto (CL) vacuum solution \citep{Campanelli-1993} to show
that the BD theory produces a wormhole when the post-Newtonian parameter
$\gamma>1$ and a naked singularity when $\gamma<1$. They also found
that the combination $(1-\gamma)(1+2\gamma)/(1+\gamma)$ determines
the sign of $G_{00}$. While the WEC is indeed violated when a wormhole
is formed $(\gamma>1)$, the reverse is not necessarily true, as a
violation can also occur when $-1<\gamma<-1/2$, in which case no
wormhole is formed.

In this paper, we revisit the analysis of ALC and delve
further into the WEC in the BD theory, aiming to answer the question
of what happens to spacetime \emph{in the absence of a wormhole} when
the WEC is violated. Our findings for BD gravity may have broader
implications for modified gravity theories at large \citep{Agnese-2001,Nandi-1997,Radu-2021,Jusufi-2022,Koutou-2020,Easson-2015,Easson-2017}.

The paper is structured as follows. Section \ref{sec:Extension-of-CL}
reviews and \emph{generalizes} the existing CL
solution so that it is also valid for the region where $0<r<r_{\text{s}}$.
This section also analyzes the singularities of the \emph{generalized}
CL solution. Section \ref{sec:Wormhole} examines a particular situation
in which the solution allows for wormholes. Section \ref{sec:Violation-WEC}
explores the violation of the WEC. Section \ref{sec:Mapping} relates
the \emph{special} Buchdahl-inspired metric uncovered and examined
in Refs.~\cite{Nguyen-2022-Lambda0,2023-WH,Nguyen-2022-Buchdahl,Buchdahl-1962}
with the \emph{generalized} CL metric. Section \ref{sec:KS} constructs
a Kruskal-Szekeres (KS) diagram for the latter metric. Appendix \ref{sec:On-Brans-solutions}
gives a brief overview of the Brans solutions, while Appendix \ref{sec:Verify}
validates the \emph{generalized} CL solution via direct inspection.

\section{\label{sec:Extension-of-CL}Extension of the Campanelli-Lousto solution}

We shall consider the original BD action \citep{BransDicke-1961}
\begin{equation}
\mathcal{S}=\int d^{4}x\sqrt{-g}\left[\phi\,\mathcal{R}-\frac{\omega}{\phi}g^{\mu\nu}\partial_{\mu}\phi\partial_{\nu}\phi\right]\label{eq:BD-action}
\end{equation}
where $\phi$ is the BD scalar field and $\omega$ is the dimensionless BD parameter. Its field equations are given by \small
\begin{align}
&G_{\mu\nu}  =\frac{1}{\phi}\left(\nabla_{\mu}\nabla_{\nu}\phi-g_{\mu\nu}\square\,\phi\right) +\frac{\omega}{\phi^{2}}\Bigl(\nabla_{\mu}\phi\nabla_{\nu}\phi-\frac{1}{2}g_{\mu\nu}(\nabla\phi)^{2}\Bigr),\nonumber\\
&(2\omega+3)\,\square\,\phi  =0.\label{eq:BD-eqn-2}
\end{align}
\normalsize where $G_{\mu\nu}=\mathcal{R}_{\mu\nu}-g_{\mu\nu}\mathcal{R}/2$ is the Einstein tensor and $\mathcal{R}$ is the Ricci scalar 
\begin{equation}
\mathcal{R}=\frac{\omega}{\phi^{2}}\,(\nabla\phi)^{2}.\label{eq:BD-Ricci}
\end{equation}

Brans discovered four solutions in~\citep{Brans-1962}, which are expressed in isotropic coordinates and named type I, II, III, and IV, respectively. Among these solutions, the Brans type I solution has been the most explored one, with the other three solutions being ``derivable'' from it via duality or by taking a proper limit, as we shall discuss in Appendix~\ref{sec:On-Brans-solutions}. In what follows, we adopt the notation in ALC~\citep{Agnese-1995}, who were the first researchers to correctly expose the wormhole and naked singularity from the CL solution \footnote{\label{fn:C-L-typos}Ref.~\citep{Agnese-1995} contains several
typos. We have identified three sets of misprint. Therein, Eq.~(8)
should be $2B+2$ in place of $2B$; in Eq.~(22) all $-B$ terms should
be $+B$; in Eq.~(24) the exponent should be $2\bigl(\sqrt{(1+\gamma)/2}-1\bigr)$ instead of $2\bigl(\sqrt{2/(1+\gamma)}-1\bigr)$.}, which is in essence the Brans type I solution expressed in a different
coordinate system \citep{Campanelli-1993,Vanzo-2012}.

The Brans type I solution comprises of a static spherisymmetric vacuum
metric \footnote{The Brans type I solution exhibits symmetry upon $\frac{4\bar{r}}{r_{\text{s}}}\leftrightarrows\frac{r_{\text{s}}}{4\bar{r}}$
producing two symmetric sheets of spacetime across the reflection
point $r_{\text{s}}/4$. }
\begin{align}
ds_{\text{Brans-I}}^{2} & =-\left|\frac{1-\frac{r_{\text{s}}}{4\bar{r}}}{1+\frac{r_{\text{s}}}{4\bar{r}}}\right|^{2A}dt^{2}\nonumber \\
& +\left(1-\frac{r_{\text{s}}^{2}}{16\bar{r}^{2}}\right)^{2}\left|\frac{1-\frac{r_{\text{s}}}{4\bar{r}}}{1+\frac{r_{\text{s}}}{4\bar{r}}}\right|^{2B}\left(d\bar{r}^{2}+\bar{r}^{2}d\Omega^{2}\right)\label{eq:Brans-I-metric}
\end{align}
and a scalar field
\begin{equation}
\phi_{\text{Brans-I}}=\phi_{0}\left|\frac{1-\frac{r_{\text{s}}}{4\bar{r}}}{1+\frac{r_{\text{s}}}{4\bar{r}}}\right|^{-(A+B)}\label{eq:Brans-I-scalar}
\end{equation}
with $\bar{r}$ being the isotropic radial coordinate. In terms of $A$ and $B$,
the BD coupling parameter is
\begin{equation}
\omega=-2\,\frac{A^{2}+AB+B^{2}-1}{(A+B)^{2}}.\label{eq:Brans-I-omega}
\end{equation}
Since $\omega$ is a parameter of the action, the Brans type I solution
effectively involves three parameters, $r_{\text{s}}$, $\phi_{0}$
and either $A$ or $B$ [related by $\omega$~\eqref{eq:Brans-I-omega}]. The
Ricci scalar is
\begin{equation}
\mathcal{R}=-(A^{2}+AB+B^{2}-1)\frac{128\,r_{\text{s}}^{-2}}{\left(\frac{4\bar{r}}{r_{\text{s}}}-\frac{r_{\text{s}}}{4\bar{r}}\right)^{4}}\left|\frac{1-\frac{r_{\text{s}}}{4\bar{r}}}{1+\frac{r_{\text{s}}}{4\bar{r}}}\right|^{-2B}.\label{eq:Brans-I-Ricci}
\end{equation}
Note that at the specific point $\{A=1,\,B=-1\}$, the metric described
by \eqref{eq:Brans-I-metric} simplifies to the Schwarzschild metric
expressed in the isotropic coordinate system and the scalar
field in \eqref{eq:Brans-I-scalar} reduces to a constant value.

A slightly more illuminating expression can be obtained by ``diagonalizing''
$A$ and $B$, namely
\begin{equation}\label{Apm}
A_{\pm}:=\frac{1}{2}(A\pm B).
\end{equation}
In this notation, the solution is given by \small
\begin{align}
& ds_{\text{Brans-I}}^{2}  =\bigg|\frac{1-\frac{r_{\text{s}}}{4\bar{r}}}{1+\frac{r_{\text{s}}}{4\bar{r}}}\bigg|^{2A_{+}}\bigg\{-\left|\frac{1-\frac{r_{\text{s}}}{4\bar{r}}}{1+\frac{r_{\text{s}}}{4\bar{r}}}\right|^{2A_{-}}dt^{2}\nonumber \\
& +\Big(1-\frac{r_{\text{s}}^{2}}{16\bar{r}^{2}}\Big)^{2}\bigg|\frac{1-\frac{r_{\text{s}}}{4\bar{r}}}{1+\frac{r_{\text{s}}}{4\bar{r}}}\bigg|^{-2A_{-}}(d\bar{r}^{2}+\bar{r}^{2}d\Omega^{2})\bigg\},\label{eq:Brans-I-metric-2}
\end{align}
\begin{equation}
\phi_{\text{Brans-I}}=\phi_{0}\bigg|\frac{1-\frac{r_{\text{s}}}{4\bar{r}}}{1+\frac{r_{\text{s}}}{4\bar{r}}}\bigg|^{-2A_{+}}.\label{eq:Brans-I-scalar-2}
\end{equation}
\normalsize The conformal factor in the metric is a reciprocal of the scalar field.
The proper part of the metric is non-Schwarzschild if $A_{-}\neq1$.
The Ricci scalar is
\begin{equation}
\mathcal{R}=\frac{256\,\omega A_{+}^{2}r_{\text{s}}^{-2}}{\left(\frac{4\bar{r}}{r_{\text{s}}}-\frac{r_{\text{s}}}{4\bar{r}}\right)^{4}}\left|\frac{1-\frac{r_{\text{s}}}{4\bar{r}}}{1+\frac{r_{\text{s}}}{4\bar{r}}}\right|^{-2A_{+}+2A_{-}}\label{eq:Brans-I-Ricci-1}
\end{equation}
The relation~\eqref{eq:Brans-I-omega} is simplified to \footnote{Note that when the dilation field is a constant, viz. $A_{+}=0$,
by virtue of \eqref{eq:A-relation}, $A_{-}=\pm1$ except for $\omega\rightarrow+\infty$.}
\begin{equation}
(2\omega+3)A_{+}^{2}+A_{-}^{2}=1.\label{eq:A-relation}
\end{equation}

\subsection{The \emph{generalized} Campanelli-Lousto solution}

From the Brans type I solution, one can obtain the CL
solution \citep{Campanelli-1993} upon making the coordinate transformation:
\begin{equation}
r=\bar{r}\left(1+\frac{r_{\text{s}}}{4\bar{r}}\right)^{2}\geqslant r_{\text{s}},
\end{equation}
or, equivalently
\begin{equation}
\left(\frac{1-\frac{r_{\text{s}}}{4\bar{r}}}{1+\frac{r_{\text{s}}}{4\bar{r}}}\right)^{2}=1-\frac{r_{\text{s}}}{r}.
\end{equation}
For each value of $r>r_{\text{s}}$ there exist two distinct values
$\bar{r}_{1}$ and $\bar{r}_{2}$ such that $\frac{4\bar{r}_{1}}{r_{\text{s}}}=\frac{r_{\text{s}}}{4\bar{r}_{2}}$,
corresponding to two symmetric exterior sheets of spacetime. In the coordinate $r$, the Brans type I solution becomes the metric \citep{Campanelli-1993}\small
\begin{equation}
ds^{2}  =-\Big(1-\frac{r_{\text{s}}}{r}\Big)^{A}dt^{2}+\Big(1-\frac{r_{\text{s}}}{r}\Big)^{B}dr^{2}+\Big(1-\frac{r_{\text{s}}}{r}\Big)^{B+1}r^{2}d\Omega^{2},\label{eq:CL-metric}
\end{equation}
\normalsize and the scalar field
\begin{align}
\phi(r) & =\phi_{0}\left(1-\frac{r_{\text{s}}}{r}\right)^{-\frac{1}{2}(A+B)},\label{eq:CL-scalar}
\end{align}
which together comprise the CL solution. It is
important to note that the above expressions, as originally reported
in \citep{Campanelli-1993}, are only applicable for the region where
$r>r_{\text{s}}$. They are not valid if $A$ or $B$ takes on a non-integer
value. However, it is a straightforward exercise to verify by \emph{direct
inspection} that the following \emph{generalized} CL metric and its associated scalar field\small
\begin{align}
& ds^{2}  =-\text{sgn}\left(1-\frac{r_{\text{s}}}{r}\right)\left|1-\frac{r_{\text{s}}}{r}\right|^{A}dt^{2}\nonumber \\
&  +\text{sgn}\left(1-\frac{r_{\text{s}}}{r}\right)\left|1-\frac{r_{\text{s}}}{r}\right|^{B}dr^{2}+\left|1-\frac{r_{\text{s}}}{r}\right|^{B+1}r^{2}d\Omega^{2},\label{eq:gen-CL-metric}
\end{align}
\begin{align}
\phi(r) & =\phi_{0}\,\text{sgn}\left(1-\frac{r_{\text{s}}}{r}\right)\left|1-\frac{r_{\text{s}}}{r}\right|^{-\frac{1}{2}(A+B)},\label{eq:gen-CL-scalar}
\end{align}
\normalsize form a solution to the vacuo field equations~\eqref{eq:BD-eqn-2}
for all values of $r\in(0,r_\text{s})\cup(r_\text{s},+\infty)$, as shown in Appendix \ref{sec:Verify}. In the above expressions, the notation ``sgn'' stands for the signum function. Obviously, the generalized CL solution \eqref{eq:gen-CL-metric}--\eqref{eq:gen-CL-scalar}
reproduces the CL solution \eqref{eq:CL-metric}--\eqref{eq:CL-scalar}
for the region where $r>r_{\text{s}}$, but it is also applicable
for the region where $0<r<r_{\text{s}}$ as well. It also recovers
the Schwarzschild metric when $\{A=1,\,B=-1\}$.

In the \emph{generalized} CL metric, $g_{tt}$ and $g_{rr}$ flip
their sign across $r=r_{\text{s}}$ as desired, and the Ricci scalar
is\small
\begin{equation}
\mathcal{R}=-(A^{2}+AB+B^{2}-1)\frac{r_{\text{s}}^{2}}{2r^{4}}\Big|1-\frac{r_{\text{s}}}{r}\Big|^{-B-2}\text{sgn}\Big(1-\frac{r_{\text{s}}}{r}\Big).\label{eq:gen-CL-Ricci}
\end{equation}
\normalsize It should be noted that although $\phi$ remains constant along the
line $A+B=0$, the value of $\omega$ becomes infinite (except at
the Schwarzschild point $\{A=1,\,B=-1\}$), as deduced from Eqs. \eqref{eq:Brans-I-scalar}
and \eqref{eq:Brans-I-omega}. This explains why the metric \eqref{eq:Brans-I-metric}
along the line $A+B=0$ can deviate from the Schwarzschild metric.


\subsection{\label{sec:Singularities}Singularities}

\begin{figure}[!t]
\begin{centering}
\includegraphics{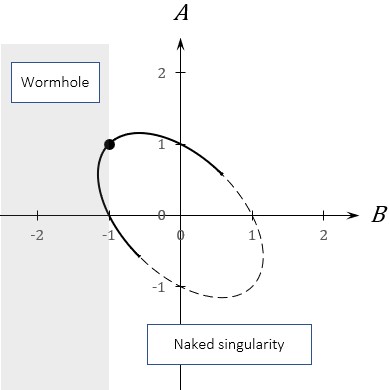}\linebreak
\par\end{centering}
\caption{\label{fig:A-vs-B}{\footnotesize Parameter space of the \emph{generalized} CL metric,
Eq. \eqref{eq:gen-CL-metric}. Black dot is Schwarzschild metric $\{A=1,B=-1\}$.
The ellipse is the loci $A^{2}+AB+B^{2}=1$ along which the Ricci
scalar $\mathcal{R}$ vanishes for all $r\in\mathbb{R}$. The vertical
line $B=-1$ separates the formation of a wormhole (the shaded area)
from the naked singularity (the white area). The solid curve segment
on the ellipse corresponds to the asymptotically flat Buchdahl-inspired
metric, discussed in Sec. \ref{sec:Mapping}, with the two end points
of the segment being ($1/\sqrt{3},\,1/\sqrt{3}$) and ($-1/\sqrt{3},\,-1/\sqrt{3}$).}}
\end{figure}

To expose the physical singularities, let us examine the Ricci scalar
and the Kretschmann invariant of the generalized CL metric. The Ricci scalar~\eqref{eq:gen-CL-Ricci} is identically zero
on the loci of an ellipse $A^{2}+AB+B^{2}=1$. See Fig. \ref{fig:A-vs-B}. The
Kretschmann invariant is given by \citep{Bronnikov-1997}
\begin{table*}[!t]
\noindent \begin{centering}
\begin{tabular}{|c|c|c|c|c|c|c|}
\hline 
$\ \ $Case$\ \ $ & Range for $B$ & $r\rightarrow0$ & $r\rightarrow r_{\text{s}}$ & $\ $$\begin{array}{c}
	\text{Wormhole or}\\
	\text{naked singularity ?}
\end{array}$$\ $ & $\begin{array}{c}
	\text{Interior region}\\
	\text{consists of ...}
\end{array}$ & $\ $$\begin{array}{c}
	\text{WEC}\\
	\text{violation ?}
\end{array}$$\ $\tabularnewline
\hline 
\hline 
{[}I{]} & $B\leqslant-2$ & $\begin{array}{c}
	R\text{ vanishes}\\
	K\text{ diverges}
\end{array}$ & $\begin{array}{c}
	R\text{ diverges}\\
	K\text{ is finite}
\end{array}$ & Wormhole & 2 unbounded sheets & Yes\tabularnewline
\hline 
{[}II{]} & $\ -2<B<-1\ $ & $\ \,\begin{array}{c}
	R\text{ vanishes}\\
	K\text{ diverges}
\end{array}\ \,$ & $\ \,\begin{array}{c}
	R\text{ diverges}\\
	K\text{ diverges}
\end{array}\ \,$ & Wormhole & $\ \ $2 unbounded sheets$\ \ $ & Yes\tabularnewline
\hline 
{[}III{]} & $-1\leqslant B\leqslant 1$ & $\begin{array}{c}
	R\text{ is finite}\\
	K\text{ diverges}
\end{array}$ & $\begin{array}{c}
	R\text{ is finite}\\
	K\text{ diverges}
\end{array}$ & Naked singularity & 4 bounded sheets & No\tabularnewline
\hline 
{[}IV{]} & $1<B<2$ & $\begin{array}{c}
	R\text{ diverges}\\
	K\text{ diverges}
\end{array}$ & $\begin{array}{c}
	R\text{ vanishes}\\
	K\text{ diverges}
\end{array}$ & Naked singularity & 2 unbounded sheets & Yes\tabularnewline
\hline 
{[}V{]} & $2\leqslant B$ & $\begin{array}{c}
	R\text{ diverges}\\
	K\text{ is finite}
\end{array}$ & $\begin{array}{c}
	R\text{ vanishes}\\
	K\text{ diverges}
\end{array}$ & Naked singularity & 2 unbounded sheets & Yes\tabularnewline
\hline 
\end{tabular}
\par\end{centering}
\caption{\label{tab:Behavior}{\footnotesize Behavior of the areal radius $R$ and the Kretschmann
scalar $K$ as $r$ approaches $0$ or $r_{\text{s}}$. A wormhole
is formed when $R$ exhibits a minimum in the region where $r>r_{\text{s}}$.
The Weak Energy Condition (WEC) is violated if $G_{00}<0$, see Section
\ref{sec:Violation-WEC}.}}
\end{table*}
\small
\begin{align}
K & :=\mathcal{R}_{\mu\nu\rho\sigma}\mathcal{R}^{\mu\nu\rho\sigma},\nonumber\\
& =4\left(\mathcal{R}_{\ \ 01}^{01}\right)^{2}+8\left(\mathcal{R}_{\ \ 02}^{02}\right)^{2}+8\left(\mathcal{R}_{\ \ 12}^{12}\right)^{2}+4\left(\mathcal{R}_{\ \ 23}^{23}\right)^{2},\nonumber\\
& =\Big|1-\frac{r_{\text{s}}}{r}\Big|^{-2(B+2)}\ \frac{r_{\text{s}}^{2}}{r^{6}}\,\Big(6\mathfrak{A}-2\mathfrak{B}\,\frac{r_{\text{s}}}{r}+\frac{\mathfrak{C}}{4}\,\frac{r_{\text{s}}^{2}}{r^{2}}\Big),\label{eq:Kretschmann-CL}
\end{align}
\normalsize in which \small
\begin{align}
\mathfrak{A} & =A^{2}+B^{2},\nonumber\\
\mathfrak{B} & =A^{2}(A-2B+3)-B(B-1)(B-2),\nonumber\\
\mathfrak{C} & =(A+1)A^{2}(A-2B+3)\nonumber \\
& \ \ \ +(3A^{2}+B^{2}-2B+3)(B-1)^{2}.
\end{align}
\normalsize The curvature singularity at $r=r_{\text{s}}$ exists
for $B>-2$, except at the two points $\{A=1,\,B=-1\}$ and $\{A=0,\,B=-1\}$.

While both the the Ricci scalar and the Kretschmann scalar vanish at $r\rightarrow+\infty$ as $\mathcal{R}\simeq r^{-4}$ and $K\simeq r^{-6}$, they project singularities at $r\rightarrow0$ or $r\rightarrow r_{\text{s}}$. In these limits, the two scalars show similar behaviors, that is $K\simeq\mathcal{R}^{2}$,
\begin{equation}
K\simeq\mathcal{R}^2\simeq\begin{cases}
\big|r-r_{\text{s}}\big|^{-2(B+2)} & \text{as }r\rightarrow r_{\text{s}}\,,\\
r^{2(B-2)} & \text{as }r\rightarrow 0\,,
\end{cases}
\end{equation}
it is thus sufficient to focus on the Kretschmann invariant and \emph{the areal radius},
$R(r):=r\left|1-\frac{r_{\text{s}}}{r}\right|^{\frac{1}{2}(B+1)}$, in our analysis. The summarized results are shown in Table~\ref{tab:Behavior}.

We want to emphasize that the two regions, $r>r_{\text{s}}$ and $r<r_{\text{s}}$,
are typically \emph{disconnected} due to the presence of physical
singularities located along the separatrix $r=r_{\text{s}}$. Generally,
there are no geodesics that can traverse from the region where $r>r_{\text{s}}$
to the region where $r<r_{\text{s}}$. However, despite their \emph{physical}
disconnection, these regions are part of the same \emph{mathematical}
solution, and their behavior can provide insights into the violation
of the WEC, which is a \emph{mathematical}
property of the solution. Table \ref{tab:Behavior} also includes
the WEC status for completeness, and an elaboration on this topic
will be presented in Section \ref{sec:Violation-WEC}.

\section{\label{sec:Wormhole}The case of $B<-1$: $\ $Wormholes in Brans-Dicke
gravity}

\begin{figure*}[!t]
\noindent \begin{centering}
\includegraphics[scale=0.86]{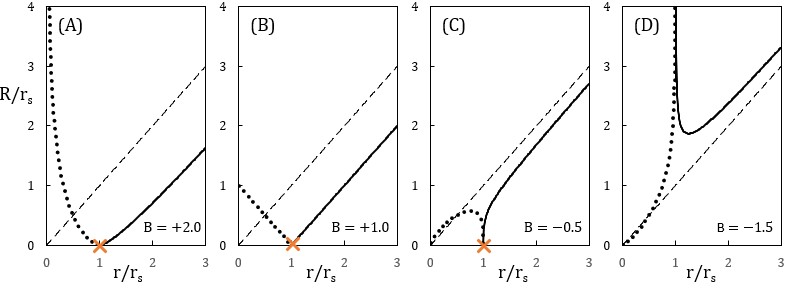}
\par\end{centering}
\caption{\label{fig:R(r)-CL}{\footnotesize Behavior of the areal radius, $R/r_{\text{s}}$,
for the \emph{generalized} CL metric versus $r/r_{\text{s}}$.
Panel D, representative of $B<-1$, yields a minimum for $R(r)$ and
corresponds to a wormhole ``throat''. Note that the horizon at $r=r_{\text{s}}$
is non-singular only if $\{B\protect\leq-2\}$ or $\{B=-1$ and $A\in\{0,1\}\}$.
Some of the plots have been extended to $r<r_{\text{s}}$ even when
the horizon is singular, shown as crosses, with the extension shown
in dotted segments. For benchmarking, the dashed lines $R=r$ are
the Schwarzschild metric, $\{B=-1$ and $A=1\}$.}}
\end{figure*}

This section revisits the analysis carried out by Agnese and La Camera
\citep{Agnese-1995}. The new element is the behavior of the areal
radius $R(r)$, produced in Fig. \ref{fig:R(r)-CL}.

\subsection{\label{subsec:MT-metric} Casting in the Morris-Thorne ansatz}

In this section, we shall bring metric \eqref{eq:gen-CL-metric} to
the Morris-Thorne ansatz \citep{MorrisThorne-1988-2}
\begin{equation}
ds^{2}=-e^{2\Phi(R)}dt^{2}+\frac{dR^{2}}{1-\frac{b(R)}{R}}+R^{2}d\Omega^{2}.
\end{equation}
As a function of $r$, the areal radius and its asymptotic behaviors are
\begin{equation}\label{eq:areal-radius}
R(r)  =r\Big|1-\frac{r_{\text{s}}}{r}\Big|^{\frac{B+1}{2}}\simeq\begin{cases}
r & \text{as }r\rightarrow\infty\\
|r-r_{\text{s}}|^{\frac{1+B}{2}} & \text{as }r\rightarrow r_{\text{s}},\\
r^{\frac{1-B}{2}} & \text{as }r\rightarrow 0
\end{cases}
\end{equation}
leading to
\begin{equation}
\frac{dR}{dr}=\text{sgn}\Big(1-\frac{r_{\text{s}}}{r}\Big)\Big|1-\frac{r_{\text{s}}}{r}\Big|^{\frac{B-1}{2}}\Big[1+\frac{B-1}{2}\frac{r_{\text{s}}}{r}\Big].\label{eq:dR-dr}
\end{equation}
The redshift function $\Phi(R)$ is defined via
\begin{equation}
e^{2\Phi(R)}=\text{sgn}\Big(1-\frac{r_{\text{s}}}{r}\Big)\Big|1-\frac{r_{\text{s}}}{r}\Big|^{A},\label{eq:redshift-func}
\end{equation}
and the shape function $b(R)$ via
\begin{equation}
1-\frac{b(R)}{R}=\frac{1}{1-\frac{r_{\text{s}}}{r}}\Big[1+\frac{B-1}{2}\frac{r_{\text{s}}}{r}\Big]^{2},\label{eq:shape-func}
\end{equation}
with $r$ being implicit function of $R$ via Eq. \eqref{eq:areal-radius}.
The behavior of $R(r)$ is shown in Fig. \ref{fig:R(r)-CL}. In the
region where $r>r_{\text{s}}$, $R(r)$ exhibits a minimum for $B<-1$
(represented by Panel D in Fig. \ref{fig:R(r)-CL}).


We must note in advance that the graphs shown in Fig. \ref{fig:R(r)-CL}
are only ``half'' of the story. There is a maximal analytic extension
of the \emph{generalized} CL metric via the KS
diagram, to be constructed in Section \ref{sec:KS}. The KS diagram
``doubles'' the coverage of the generalized CL solution, in addition
to uncovering a ``gulf'' sandwiched between the four quadrants.

\subsection{\label{subsec:MT-constraints}The four Morris-Thorne constraints}

Let us restrict our consideration to the range $B<-1$, of which Panel
D in Fig. \ref{fig:R(r)-CL} is a representative. 
\begin{itemize}
\item The areal radius $R$ diverges at $r=r_{\text{s}}$ when $B<-1$.
\item The equation $\frac{dR}{dr}=0$ in \eqref{eq:dR-dr} has a single
root
\begin{equation}
r_{*}=\frac{1-B}{2}\,r_{\text{s}}>r_{\text{s}}\ \ \ \text{when }B<-1,\label{eq:r-star}
\end{equation}
with $R$ attaining a \emph{minimum} value
\begin{align}
R_{*} & =\frac{r_{\text{s}}}{2}\sqrt{B^{2}-1}~\left(\frac{B+1}{B-1}\right)^{\frac{B}{2}}.\label{eq:R-star}
\end{align}
\end{itemize}
It is straightforward to verify that when $B<-1$ the four constraints
for the metric to possess a wormhole are met \cite{MorrisThorne-1988-1,MorrisThorne-1988-2},
as can be seen in Panel D of Fig. \ref{fig:R(r)-CL}: (i) The redshift function $\Phi(R)$ [defined in \eqref{eq:redshift-func}] is finite everywhere (hence no horizon); (ii) At the throat of the wormhole, $R_{*}$ is the minimum value of
$R$ per Eq. \eqref{eq:R-star}; (iii) The proper radial distance is finite, i.e. $b(R)/R\leq1$ for $R\geq R_{*}$. The equality sign holds only
at the throat, $R=R_{*}$; (iv) $ \lim_{R\rightarrow+\infty}[b(R)/R]=0$ to ensure asymptotic flatness. We emphasize that the throat occurs at location $r_*$ given in Eq.~\eqref{eq:r-star}, where the scalars $\mathcal{R}$ and $K$ do not diverge (viz. Case I and Case II in Table~\ref{tab:Behavior}), and that, to ensure regularity of the solution everywhere, the lower sheet of the wormhole is a symmetric copy of the existing one (the one corresponding to $r\geq r_{*}$). The two sheets are \emph{smoothly} glued together at $r_*$, ensuring the two-way traversability across the throat from one sheet to the other~\cite{2023-WH}. The wormhole thus does \emph{not} include the spacetime points where the two scalars diverge.

\section{\label{sec:Violation-WEC}Violation of the Weak Energy Condition}

In its formal geometric form \cite{Koutou-2020}, the WEC~\eqref{WEC} is violated
if the energy density is negative in some region.

The generalized CL metric given in \eqref{eq:gen-CL-metric} has the
$00-$component of the Einstein tensor
\begin{equation}
G_{00}=\frac{1-B^{2}}{4\,r^{4}}r_{\text{s}}^{2}\Big|1-\frac{r_{\text{s}}}{r}\Big|^{A-B-2}.\label{eq:G00}
\end{equation}
Regardless of the value of $A$, $G_{00}<0$ when a ``throat'' is
formed, viz. when $B<-1$; see Panel D in Fig. \ref{fig:R(r)-CL}.
The inequality $G_{00}<0$ is interpreted as a signature of negative
energy density, resulted from the BD scalar field. 

The appearance of a wormhole
is associated with a violation of the WEC. However, not all violations
of the WEC lead to a wormhole. For example, when $B>1$, the WEC is
violated [$G_{00}<0\ \ \forall r$ as shown in Eq. \eqref{eq:G00}],
but a wormhole does not appear. This raises the question of what other
effects a violation of the WEC can have.

Figure~\ref{fig:R(r)-CL} offers an answer to this question. Panels
A and D both violate the WEC, but only Panel D exhibits a wormhole,
while Panel A does not. However, both panels have something in common:
they both feature an unbounded sheet of spacetime in the region where
$0<r<r_{\text{s}}$ that could exist independently of the region where
$r>r_{\text{s}}$. In contrast, Panel C has a confined domain consisting
of two finite-size ``bubbles'' glued together \footnote{Note that the region where $0<r<r_{\text{s}}$ has a mirror image in
the KS diagram; see Section \ref{sec:KS}.}. Therefore, a violation of the WEC can alter the topology of spacetime,
including that of the region $0<r<r_{\text{s}}$. 

For the generalized CL metric, a violation of the WEC leads to a divergence
in the cross-section area of the spacetime configuration. This observation
might have a broader range of applicability in wormhole physics, beyond
Brans-Dicke gravity.

The existence of a wormhole is not a direct consequence of a violation
of the WEC, but rather a by-product of a divergence in the areal radius
$R(r)$ at a finite value of $r$ (whether at $r=0$ or $r=r_{\text{s}}$).
Thus, a violation of the WEC may or may not result in a wormhole,
and only when the divergence of $R(r)$ occurs at $r_{\text{s}}$
does a wormhole form.

\section{\label{sec:Mapping}Mapping the asymptotically flat Buchdahl-inspired
metric into the generalized Campanelli-Lousto metric}

A particularly interesting case is the loci $A^{2}+AB+B^{2}=1$ in
Fig. \ref{fig:A-vs-B}, where $\omega=0$ throughout except at ${A=+1,B=-1}$.
This loci corresponds to a BD theory with $\omega=0$.

In this section, we aim to establish a connection between the \emph{generalized}
CL solution and a vacuum solution that is asymptotically flat at spatial
infinity, discovered for pure $\mathcal{R}^{2}$ gravity in~\citep{Nguyen-2022-Lambda0} (see also~\cite{Nguyen-2022-Buchdahl}):
\begin{equation}
ds^{2}  =|f(r)|^{\tilde{k}}\Big[ -f(r)dt^{2}+\frac{dr^{2}}{f(r)}\frac{\rho^{4}(r)}{r^{4}}+\rho^{2}(r)d\Omega^{2}\Big], \label{eq:special-B-1}
\end{equation}
with
\begin{align}
\rho(r) & =\zeta r_{\text{s}}\frac{|f(r)|^{\frac{\zeta-1}{2}}}{1-\text{sgn}(f(r))|f(r)|^{\zeta}},\label{eq:special-B-2}\\
f(r) & := 1-\frac{r_{\text{s}}}{r}\,,\quad \zeta :=\sqrt{1+3\tilde{k}^{2}}\,,\quad \tilde{k}:=\frac{k}{r_{\text{s}}},\label{eq:special-B-3}
\end{align}
which we named the \emph{special} Buchdahl-inspired metric to reflect
its distinctiveness among the class of Buchdahl-inspired metrics.
This \emph{special}
metric has a vanishing Ricci scalar everywhere, $\mathcal{R}\equiv 0\ \forall r$, and it is asymptotically flat but not Ricci flat: $\mathcal{R}_{\mu\nu}\neq 0$~\cite{Nguyen-2022-Lambda0}. It is worth mentioning that the discussion made by Higgs~\cite{Higgs}, following his Eq.~(9), concerns only Ricci flat solutions since if $\lambda =0$, $R_{\mu\nu}=0$ too.

To connect the \emph{special} Buchdahl-inspired metric with the \emph{generalized}
CL metric, we introduce a new radial coordinate $r'$ such that
\begin{equation}
1-\frac{\zeta r_{\text{s}}}{r'}:=\text{sgn}\Big(1-\frac{r_{\text{s}}}{r}\Big)\Big|1-\frac{r_{\text{s}}}{r}\Big|^{\zeta}
\end{equation}
The metric presented in \eqref{eq:special-B-1}--\eqref{eq:special-B-3}
can be transformed into 
\begin{align}
ds^{2} & =-\Big|1-\frac{\zeta r_{\text{s}}}{r'}\Big|^{\frac{\tilde{k}+1}{\zeta}-1}\Big(1-\frac{\zeta r_{\text{s}}}{r'}\Big)dt^{2}\nonumber \\
& \ \ \ +\Big|1-\frac{\zeta r_{\text{s}}}{r'}\Big|^{\frac{\tilde{k}-1}{\zeta}+1}\Big[\frac{dr'^{2}}{1-\frac{\zeta r_{\text{s}}}{r'}}+r'^{2}d\Omega^{2}\Big],
\end{align}
which is precisely the \emph{generalized} CL metric given by
Eq. \eqref{eq:gen-CL-metric}, with an ``effective'' Schwarzschild
radius $\zeta\,r_{\text{s}}$, where
\begin{equation}
A=\frac{\tilde{k}+1}{\zeta};\ \ \ B=\frac{\tilde{k}-1}{\zeta}.\label{eq:A-B-special}
\end{equation}
It is straightforward to check that the parameters defined in \eqref{eq:A-B-special}
obey the equality (given Eq. \ref{eq:special-B-3})
\begin{equation}
A^{2}+AB+B^{2}=1
\end{equation}
Therefore, the \emph{special} Buchdahl-inspired metric is identical
with the \emph{generalized} CL metric with $\omega=0$.

The \emph{special} Buchdahl-inspired metric \emph{simultaneously}
belongs to the general Buchdahl-inspired metric family when $\Lambda=0$~\cite{Nguyen-2022-Lambda0,Nguyen-2022-Buchdahl}
and to the \emph{generalized} CL solution family when $\omega=0$,
representing the intersection of the two families. As such, it is
a vacuum solution to two theories at the same time. With $A$ and $B$ identified as in \eqref{eq:A-B-special}, the \emph{special}
Buchdahl-inspired metric is represented by a solid curve segment in
Fig.$\ $\ref{fig:R(r)-CL}, which corresponds to a portion of the
ellipse defined by $A^{2}+AB+B^{2}=1$.

\section{\label{sec:KS}Constructing Kruskal-Szekeres diagram for the generalized
Campanelli-Lousto metric}

\begin{figure}[!t]
\noindent \begin{centering}
\hskip-6pt\includegraphics[scale=0.52]{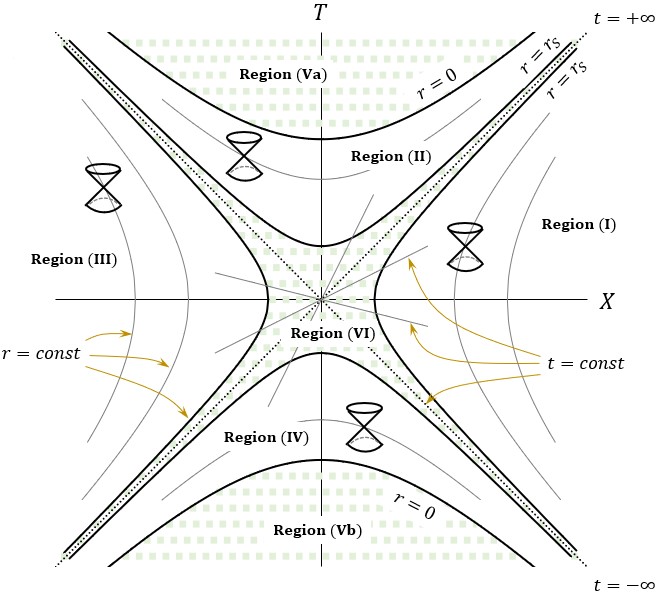}
\par\end{centering}
\caption{\label{fig:KS-diagram}{\footnotesize Kruskal-Szekeres diagram for the generalized
CL metric when $A_{-}\protect\neq1$. The \textquotedblleft gulf\textquotedblright{}
shown as Region (VI) generally involves curvature singularity at $r=r_{\text{s}}$.}}
\end{figure}

In our previous work \citep{Nguyen-2022-Lambda0}, we constructed
the KS diagram for the \emph{special} Buchdahl-inspired
vacuum solution in pure $\mathcal{R}^{2}$ gravity. As demonstrated
in the preceding section, this solution has an intimate relationship
with the \emph{generalized} CL metric. Therefore, we can adapt the
construction method from Ref.~\citep{Nguyen-2022-Lambda0} to
create a KS diagram for the \emph{generalized} CL metric,
with suitable adjustments. Figure \ref{fig:KS-diagram} shows the
resulting diagram, which we present in this section.

\subsection{\label{subsec:KS-coord}The Kruskal-Szekeres coordinates}

The tortoise coordinate $r^{*}(r)$ for the generalized CL metric
is defined to satisfy
\begin{equation}
dr^{*}=\frac{\text{sgn}\big(1-\frac{r_{\text{s}}}{r}\big)}{\big|1-\frac{r_{\text{s}}}{r}\big|^{A_{-}}}\,dr.\label{eq:tortois-eqn}
\end{equation}
The tortoise coordinate (modulo an additive constant) involves a Gaussian
hypergeometric function~\cite{Nguyen-2022-Lambda0}:\small
\begin{equation*}
r^{*}  =\frac{r_{\text{s}}}{1-A_{-}}\,\Big|1-\frac{r_{\text{s}}}{r}\Big|^{1-A_{-}}\,_{2}F_{1}\Big(2,1-A_{-};2-A_{-};1-\frac{r_{\text{s}}}{r}\Big).
\end{equation*}
\normalsize Furthermore, by integrating Eq. \eqref{eq:tortois-eqn},
the difference [if $A_{-}\in(-1,1)$]
\begin{align}
r^{*}|_{r=0}-r^{*}|_{r=r_{\text{s}}} & =\int_{0}^{r_{\text{s}}}\frac{dr}{\big(\frac{r_{\text{s}}}{r}-1\big)^{A_{-}}}=\frac{\pi A_{-}r_{\text{s}}}{\sin\pi A_{-}}.
\end{align}
We shall choose the additive constant such that the tortoise coordinate
vanishes at $r=0$. 

The advanced and retarded Eddington-Finkelstein coordinates are defined
as \citep{Eddington-1924,Finkelstein-1958}
\begin{equation}
v  :=t+r^{*}\,,\quad u  :=t-r^{*}\,,
\end{equation}
which, together with \eqref{eq:tortois-eqn}, give
\begin{equation}
du\,dv=dt^{2}-dr^{*2}=dt^{2}-\frac{dr^{2}}{\big|1-\frac{r_{\text{s}}}{r}\big|^{2A_{-}}}.
\end{equation}
Metric \eqref{eq:gen-CL-metric} then becomes
\begin{align}
ds^{2} & =\Big|1-\frac{r_{\text{s}}}{r}\Big|^{A_{+}}\biggl\{-\text{sgn}\Big(1-\frac{r_{\text{s}}}{r}\Big)\Big|1-\frac{r_{\text{s}}}{r}\Big|^{A_{-}}du\,dv\nonumber \\
& +r^{2}\Big|1-\frac{r_{\text{s}}}{r}\Big|^{-A_{-}+1}d\Omega^{2}\biggr\}.\label{eq:metric-dudv}
\end{align}

For the KS coordinates \citep{Kruskal-1960,Szekeres-1960},
we shall consider the two regions $r>r_{\text{s}}$ and $r<r_{\text{s}}$
separately.

\paragraph{The $r>r_{\text{s}}$ region:}

Let us define
\begin{equation}
X  :=\frac{1}{2}\left(e^{\frac{v}{2r_{\text{s}}}}+e^{-\frac{u}{2r_{\text{s}}}}\right),\; 
T  :=\frac{1}{2}\left(e^{\frac{v}{2r_{\text{s}}}}-e^{-\frac{u}{2r_{\text{s}}}}\right),
\end{equation}
giving
\begin{equation}
dT^{2}-dX^{2}=\frac{e^{\frac{r^{*}}{r_{\text{s}}}}}{4r_{\text{s}}^{2}}\left[dt^{2}-\frac{dr^{2}}{\left|1-\frac{r_{\text{s}}}{r}\right|^{2A_{-}}}\right]=\frac{e^{\frac{r^{*}}{r_{\text{s}}}}}{4r_{\text{s}}^{2}}dudv\,.
\end{equation}
Metric \eqref{eq:metric-dudv} becomes
\begin{align}
ds^{2} & =\left|1-\frac{r_{\text{s}}}{r}\right|^{A_{+}}\times\biggl\{-4r_{\text{s}}^{2}e^{-\frac{r^{*}}{r_{\text{s}}}}\left|1-\frac{r_{\text{s}}}{r}\right|^{A_{-}}\left(dT^{2}-dX^{2}\right)\nonumber \\
& +r^{2}\left|1-\frac{r_{\text{s}}}{r}\right|^{-A_{-}+1}d\Omega^{2}\biggr\}.
\end{align}
\normalsize

\paragraph{The $r<r_{\text{s}}$ region:}

Let us define
\begin{equation}
X  :=\frac{1}{2}\left(e^{\frac{v}{2r_{\text{s}}}}-e^{-\frac{u}{2r_{\text{s}}}}\right),\;
T  :=\frac{1}{2}\left(e^{\frac{v}{2r_{\text{s}}}}+e^{-\frac{u}{2r_{\text{s}}}}\right),
\end{equation}
giving
\begin{equation}
dT^{2}-dX^{2}=-\frac{e^{\frac{r^{*}}{r_{\text{s}}}}}{4r_{\text{s}}^{2}}\left[dt^{2}-\frac{dr^{2}}{\left|1-\frac{r_{\text{s}}}{r}\right|^{2A_{-}}}\right]=-\frac{e^{\frac{r^{*}}{r_{\text{s}}}}}{4r_{\text{s}}^{2}}dudv\,.
\end{equation}
Metric \eqref{eq:metric-dudv} becomes
\begin{align}
ds^{2} & =\left|1-\frac{r_{\text{s}}}{r}\right|^{A_{+}}\times\biggl\{-4r_{\text{s}}^{2}e^{-\frac{r^{*}}{r_{\text{s}}}}\left|1-\frac{r_{\text{s}}}{r}\right|^{A_{-}}\left(dT^{2}-dX^{2}\right)\nonumber \\
& +r^{2}\left|1-\frac{r_{\text{s}}}{r}\right|^{-A_{-}+1}d\Omega^{2}\biggr\}.
\end{align}
\normalsize

\paragraph*{In combination:}

The generalized CL metric in the KS coordinates
is thus \small
\begin{align}
ds^{2} & =\Big|1-\frac{r_{\text{s}}}{r}\Big|^{A_{+}}\times\biggl\{-4r_{\text{s}}^{2}e^{-\frac{r^{*}}{r_{\text{s}}}}\Big|1-\frac{r_{\text{s}}}{r}\Big|^{A_{-}}\Big(dT^{2}-dX^{2}\Big)\nonumber \\
& +r^{2}\Big|1-\frac{r_{\text{s}}}{r}\Big|^{-A_{-}+1}d\Omega^{2}\biggr\},\label{eq:KS-metric}
\end{align}
\normalsize
\begin{align}
T^{2}-X^{2} & =-\text{sgn\ensuremath{\Big(\ensuremath{1-\frac{r_{\text{s}}}{r}}\Big)}}e^{\frac{r^{*}}{r_{\text{s}}}},\label{eq:T-X-1}\\
\frac{T}{X} & =\Big(\tanh\frac{t}{2r_{\text{s}}}\Big)^{\text{sgn\ensuremath{\Big(1-\frac{r_{\text{s}}}{r}\Big)}}}.\label{eq:T-X-2}
\end{align}

\subsection{\label{subsec:Aspects}Key aspects of the Kruskal-Szekeres diagram
of the generalized Campanelli-Lousto metric}

Along the radial direction, $d\theta=d\phi=0$, metric \eqref{eq:KS-metric}
is
\begin{equation}
ds^{2}=-4r_{\text{s}}^{2}e^{-\frac{r^{*}}{r_{\text{s}}}}\left|1-\frac{r_{\text{s}}}{r}\right|^{A_{+}+A_{-}}\left(dT^{2}-dX^{2}\right).\label{eq:zeta-KS}
\end{equation}
The KS plane for metric \eqref{eq:zeta-KS} (to
be called \emph{the CL--KS diagram} hereafter) is illustrated in
Fig. \ref{fig:KS-diagram}, which exhibits several significant characteristics:
\begin{itemize}
\item 
The CL--KS
diagram is \emph{conformally Minkowski}. The null geodesics are $dX=\pm dT$.
Therefore, light cones align along the $45^{\circ}$ lines in the
CL--KS plane. 
\item The CL--KS diagram essentially preserves the causal structure of
the standard KS diagram, albeit with some quantitative modifications.
A constant--$r$ contour corresponds to a hyperbola, while a constant--$t$
contour corresponds to a straight line running through the origin
of the $(T,X)$ plane. The coordinate origin $r=0$ amounts to $T^{2}-X^{2}=1$,
since $r^{*}(r=0)=0$.
\item The separatrix $r=r_{\text{s}}$ is represented by \emph{two} distinct
hyperbolae, per
\begin{equation}
T^{2}-X^{2}=\begin{cases}
-e^{-\frac{\pi A_{-}}{\sin\pi A_{-}}} & \text{for }r>r_{\text{s}}\\
+e^{-\frac{\pi A_{-}}{\sin\pi A_{-}}} & \text{for }r<r_{\text{s}}
\end{cases}\label{eq:hyperbolae}
.
\end{equation}
It is worth noting that each hyperbola has two separate branches on
its own. For $A_{-}=1$, the hyperbolae \eqref{eq:hyperbolae} degenerate
into two straight lines, $T=\pm X$, as is expected for Schwarzschild
black holes. In this limit, Region (VI) disappears.
\item Region (I) extends up to the right branch of the hyperbola $T^{2}-X^{2}=-e^{-\frac{\pi A_{-}}{\sin\pi A_{-}}}$.
Region (II) extends up to the upper branch of the hyperbola $T^{2}-X^{2}=+e^{-\frac{\pi A_{-}}{\sin\pi A_{-}}}$.
\item Regions (III) and (IV) are mirror images of Regions (I) and (II),
upon flipping the sign of the KS coordinates, viz. $(T,X)\leftrightarrow(-T,-X)$.
Regions (Va) and (Vb) are unphysical, viz. $r<0$.
\item Region (VI) generally contains the curvature singularity (encoded
in the Kretschmann invariant per Eq. \eqref{eq:Kretschmann-CL}) on
the separatrix $r=r_{\text{s}}$. The region sandwiches between the
four hyperbola branches given by \eqref{eq:hyperbolae} and disappears
when $A_{-}=1$.
\end{itemize}
Further discussions about the causal structure of the $\zeta-$KS
diagram (for the \emph{special} Buchdahl-inspired metric) have been
made in Ref.$\ $\citep{Nguyen-2022-Lambda0}, and we shall not reiterate
them here. These discussions are equally applicable for the present
case, e.g. the CL--KS diagram, with Region (VI) being a new feature
compared with the KS diagram of the Schwarzschild metric.

In summary, the CL--KS diagram represents the maximal analytic extension
of the generalized CL metric. The emergence of the ``gulf'' in the
CL--KS diagram indicates fundamental changes in the topology of spacetime
around a mass source when the metric parameters deviate from their
Schwarzschild value, that is when $(A_{+},A_{-})\neq(0,1)$.

\section{\label{sec:Conclusions}Conclusions}

Our paper took a crucial step forward by generalizing the CL solution presented in Ref.$\ $\citep{Campanelli-1993}, which
was only valid for the region where $r>r_{\text{s}}$. Our new \emph{generalized}
CL solution holds for all values of the radial coordinate $r\in\mathbb{R}^{+}$,
making it applicable to both regions, $r>r_{\text{s}}$ and $0<r<r_{\text{s}}$.
We stress that although in general the two regions are \emph{physically
disconnected} by a loci of physical singularities at $r=r_{\text{s}}$,
they are both part of the same \emph{mathematical} solution.  Equipped
with this generalization, we are able to revisit and expand upon the
analysis previously produced by ALC~\citep{Agnese-1995},
gaining new insights.

The \emph{generalized} CL metric depends on four parameters: the Schwarzschild
radius $r_{\text{s}}$, the asymptotic value of the scalar field at
spatial infinity $\phi_{0}$, and the exponents $A$ and $B$ of the
metric components $g_{tt}$ and $g_{rr}$, respectively. The values
of $A$ and $B$ are related by the BD parameter $\omega$.
By examining the metric on a two-dimensional plane $(A,B)$, we found
that for $B$ in the range $(-\infty,-1)$, the metric supports a
Morris-Thorne wormhole, while for $B$ in the range $(-1,+\infty)$,
a naked singularity results. The value of $A$ has no effect on this
behavior.

We further discovered that violating the WEC
does not necessarily lead to the formation of a wormhole. However,
it does result in a divergence of the areal radius in the region where
$0<r<r_{\text{s}}$, occurring at either $r=0$ or $r=r_{\text{s}}$.
This means that WEC violation causes a change in the topology of spacetime,
including in the region where $0<r<r_{\text{s}}$. A wormhole is formed
only if the divergence takes place at $r=r_{\text{s}}$. Therefore,
a wormhole is only an indirect consequence of WEC violation and a
by-product of this topology change.

Finally, we established a connection between the \emph{special} Buchdahl-inspired
metric, that is known to be asymptotically flat for pure $\mathcal{R}^{2}$
gravity \citep{Nguyen-2022-Lambda0}, and the generalized CL metric.
This connection guided us to construct the maximal analytic extension
of the generalized CL metric. Figure \ref{fig:KS-diagram} depicts
the KS diagram for this metric, with Region (VI) that
sandwiches between the four quadrants representing a new feature for
Brans-Dicke gravity.

Overall, our findings indicate a complex interplay between the WEC
violation and wormhole formation in Brans-Dicke gravity. The novel
features observed in our study provide new insights into the behavior
of gravity in this theory, and may have implications for modified
theories of gravitation at large.
\begin{acknowledgments}
We thank the anonymous referees for their valuable feedbacks which helped improve the exposition of this paper. HKN thanks Tiberiu Harko for his helpful insight in conceptualizing
this research. We thank Carlos O. Lousto and Valerio Faraoni for their
comments, and Valerio Faraoni for pointing out the relevant Refs.$\,$\citep{Faraoni-2018,Bronnikov-1973}.
\end{acknowledgments}

\appendix

\section{\label{sec:On-Brans-solutions}ON BRANS SOLUTIONS OF TYPE II,
III AND IV}

Although the relations among the four types of the Brans solutions
have been covered in \citep{Sarkar-Bhadra-2006,Faraoni-2016,Faraoni-2018,Bronnikov-1973},
we shall reveal one more relation which has been obscure.\linebreak{}

\vskip-4pt The Brans type IV solution can be obtained from the Brans
type I solution by sending $A$ and $B$ to infinity and $r_{\text{s}}$
to zero while keeping the products $A\,r_{\text{s}}$ and $B\,r_{\text{s}}$
fixed. This can be seen as follows. When $r_{\text{s}}\rightarrow0$,
the terms $1\pm\frac{r_{\text{s}}}{4\bar{r}}$ are approximately $e^{\pm\frac{r_{\text{s}}}{4\bar{r}}}$;
thus the Brans type I metric \eqref{eq:Brans-I-metric} and scalar
field \eqref{eq:Brans-I-scalar} respectively become
\begin{align}
ds^{2} & \approx-e^{-\frac{Ar_{\text{s}}}{\bar{r}}}dt^{2}+e^{-\frac{Br_{\text{s}}}{\bar{r}}}\left(d\bar{r}^{2}+\bar{r}^{2}d\Omega^{2}\right)\\
\phi & \approx\phi_{0}\,e^{\frac{(A+B)r_{\text{s}}}{2\bar{r}}}
\end{align}
Denote $B=-(C+1)\,A$ and $r'_{\text{s}}=A\,r_{\text{s}}$. These
expressions yield
\begin{align}
ds_{\text{Brans-IV}}^{2} & =-e^{-\frac{r'_{\text{s}}}{\bar{r}}}dt^{2}+e^{\frac{(C+1)r'_{\text{s}}}{\bar{r}}}\left(d\bar{r}^{2}+\bar{r}^{2}d\Omega^{2}\right)\\
\phi_{\text{Brans-IV}} & =\phi_{0}\,e^{-\frac{Cr'_{\text{s}}}{2\bar{r}}}
\end{align}
which form the Brans type IV solution. For a given value of $\omega\in\mathbb{R}$,
the relation in \eqref{eq:Brans-I-omega} reads
\begin{equation}
A^{2}-(C+1)A^{2}+(C+1)^{2}A^{2}-1=-\frac{\omega}{2}C^{2}A^{2}
\end{equation}
With $A$ being sent to infinity, this relation yields
\begin{equation}
(\omega+2)C^{2}+2C+2=0
\end{equation}
hence
\begin{equation}
C_{\pm}=\frac{-1\pm\sqrt{-(2\omega+3)}}{\omega+2}
\end{equation}

The Brans type III solution is trivially the ``mirror'' image of
Type IV by a reflection, $\frac{\bar{r}}{r'_{\text{s}}}\leftrightarrows\frac{r'_{\text{s}}}{\bar{r}}$
. That is
\begin{align}
ds_{\text{Brans-III}}^{2} & =-e^{-\frac{\bar{r}}{r'_{\text{s}}}}dt^{2}+\frac{r_{\text{s}}^{4}}{\bar{r}^{4}}e^{\frac{(C+1)\bar{r}}{r_{'\text{s}}}}\left(d\bar{r}^{2}+\bar{r}^{2}d\Omega^{2}\right)\\
\phi_{\text{Brans-III}} & =\phi_{0}\,e^{-\frac{C\bar{r}}{2r_{s}}}
\end{align}

As was noticed in \citep{Sarkar-Bhadra-2006}, the Brans type II solution
is a curious case. It is a ``continuation into the complex plane''
by making a formal replacement in the Brans type I solution \eqref{eq:Brans-I-metric}--\eqref{eq:Brans-I-scalar},
per
\begin{equation}
r_{\text{s}}\rightarrow i\,r_{\text{s}};\ \ \ A\rightarrow i\,A;\ \ \ B\rightarrow i\,B
\end{equation}
That is
\begin{align}
ds_{\text{Brans-II}}^{2} & =-e^{4A\arctan\frac{r_{\text{s}}}{4\bar{r}}}dt^{2}\nonumber \\
& +\left(1+\frac{r_{\text{s}}^{2}}{16\bar{r}^{2}}\right)^{2}e^{4B\arctan\frac{r_{\text{s}}}{4\bar{r}}}\left(d\bar{r}^{2}+\bar{r}^{2}d\Omega^{2}\right)\\
\phi_{\text{Brans-II}} & =\phi_{0}\,e^{-2(A+B)\arctan\frac{r_{\text{s}}}{4\bar{r}}}
\end{align}
The relation in \eqref{eq:Brans-I-omega} becomes
\begin{equation}
\omega_{\text{Brans-II}}=-2\,\frac{A^{2}+AB+B^{2}+1}{(A+B)^{2}}\in\mathbb{R^{-}}
\end{equation}

\section{DIRECT VERIFICATION OF THE \emph{GENERALIZED} CAMPANELLI-LOUSTO
SOLUTION\label{sec:Verify}}

We directly check that the \emph{generalized}
CL metric and scalar field, given in Eqs. \eqref{eq:gen-CL-metric}
and \eqref{eq:gen-CL-scalar}, constitute a vacuo solution to the
BD field equations. We set $\nabla_{\mu}\phi\nabla_{\nu}\phi-g_{\mu\nu}\nabla_{\lambda}\phi\nabla^{\lambda}\phi/2=X_{\mu\nu}$
in BD field equations in vacuo~\eqref{eq:BD-eqn-2}. With Eqs. \eqref{eq:gen-CL-metric} and \eqref{eq:gen-CL-scalar},
$G_{\mu\nu}$ reads\small
\begin{align*}
G_{00} & =\frac{1-B^{2}}{4r^{4}}r_{\text{s}}^{2}\left|1-\frac{r_{\text{s}}}{r}\right|^{A-B-2}\\
G_{11} & =\frac{r_{\text{s}}}{4r^{4}}\frac{(B^{2}+2AB-2(A+B)+1)r_{\text{s}}+4(A+B)r}{\left(1-\frac{r_{\text{s}}}{r}\right)^{2}}\\
G_{22} & =\frac{r_{\text{s}}}{4r^{2}}\frac{(A^{2}+A+B-1)r_{\text{s}}-2(A+B)r}{1-\frac{r_{\text{s}}}{r}}\\
G_{33} & =G_{22}\,\sin^{2}\theta .
\end{align*}
\normalsize Next, we have
\begin{align*}
\sqrt{-g} & =\left|1-\frac{r_{\text{s}}}{r}\right|^{\frac{A}{2}+\frac{3B}{2}+1}r^{2}\sin\theta
\end{align*}
\begin{align}
\sqrt{-g}\,\square\,\phi 
& =\sin\theta\,\partial_{r}\Big\{\Big|1-\frac{r_{\text{s}}}{r}\Big|^{\frac{A}{2}+\frac{B}{2}}\Big(1-\frac{r_{\text{s}}}{r}\Big)r^{2}\times\nonumber \\
& \ \ \ \ \ \partial_{r}\Big[\phi_{0}\,\text{sgn}\left(1-\frac{r_{\text{s}}}{r}\right)\left|1-\frac{r_{\text{s}}}{r}\right|^{-\frac{A+B}{2}}\Big]\Big\}\\
& =0\text{\ \ \ \ \ }\forall r\neq r_{\text{s}}.\nonumber
\end{align}
We also have\small
\begin{align}
\frac{1}{\phi}\nabla_{0}\nabla_{0}\phi 
& =\frac{\phi_{0}r_{\text{s}}^{2}}{4r^{4}}\Big|1-\frac{r_{\text{s}}}{r}\Big|^{A-B-2}A(A+B)\\
\frac{1}{\phi}\nabla_{1}\nabla_{1}\phi 
& =\frac{\phi_{0}r_{\text{s}}}{4r^{4}}\Big|1-\frac{r_{\text{s}}}{r}\Big|^{-2}\times\\
& \Big[(2B^{2}+3AB-2(A+B)+A^{2})r_{\text{s}}+4(A+B)r\Big]\nonumber\\
\frac{1}{\phi}\nabla_{2}\nabla_{2}\phi 
& =-\frac{\phi_{0}r_{\text{s}}}{4r^{2}}\Big(1-\frac{r_{\text{s}}}{r}\Big)^{-1}\times\\
& \Big[(B^{2}+AB-(A+B))r_{\text{s}}+2(A+B)r\Big]\nonumber\\
\frac{1}{\phi}\nabla_{3}\nabla_{3}\phi & =\frac{1}{\phi}\nabla_{2}\nabla_{2}\phi\,\sin^{2}\theta
\end{align}
\normalsize and\small
\begin{align}
X_{00} 
& =\phi_{0}^{2}r_{\text{s}}^{2}\frac{(A+B)^{2}}{8r^{4}}\Big|1-\frac{r_{\text{s}}}{r}\Big|^{-2B-2}\\
X_{11} 
& =\phi_{0}^{2}r_{\text{s}}^{2}\frac{(A+B)^{2}}{8r^{4}}\Big|1-\frac{r_{\text{s}}}{r}\Big|^{-A-B-2}\\
X_{22} 
& =-\phi_{0}^{2}r_{\text{s}}^{2}\frac{(A+B)^{2}}{8r^{2}\Big(1-\frac{r_{\text{s}}}{r}\Big)}\Big|1-\frac{r_{\text{s}}}{r}\Big|^{-A-B}\\
X_{33} & =X_{22}\,\sin^{2}\theta
\end{align}
\normalsize From this stage, it is straightforward to verify, components by components, that the \emph{generalized} CL metric and scalar field in Eqs.~\eqref{eq:gen-CL-metric} and \eqref{eq:gen-CL-scalar} satisfy the BD field equations, viz. Eqs.~\eqref{eq:BD-eqn-2} ($\square\,\phi=0$), for all values of $r\in(0,r_\text{s})\cup(r_\text{s},+\infty)$.

\end{document}